\documentclass[]{pasj01}

\begin{document} 
\Received{2015/3/25}
\Accepted{2015/4/16}

\title{High-resolution ALMA observations of SDP.81. I.\ The innermost mass profile of the lensing elliptical galaxy probed by 30 milli-arcsecond images}

\author{Yoichi~\textsc{Tamura}\altaffilmark{1}%
}\email{ytamura@ioa.s.u-tokyo.ac.jp}

\author{Masamune~\textsc{Oguri},\altaffilmark{2,3,4}}

\author{Daisuke~\textsc{Iono}\altaffilmark{5,6}}

\author{Bunyo~\textsc{Hatsukade}\altaffilmark{5}}

\author{Yuichi~\textsc{Matsuda}\altaffilmark{5,6}}

\author{Masao~\textsc{Hayashi}\altaffilmark{5}}

\altaffiltext{1}{Institute of Astronomy, University of Tokyo, Osawa, Mitaka, Tokyo 181-0015}
\altaffiltext{2}{Research Center for the Early Universe, University of Tokyo, Hongo, Bunkyo-ku, Tokyo 113-0033}
\altaffiltext{3}{Department of Physics, University of Tokyo, Hongo, Bunkyo-ku, Tokyo 113-0033}
\altaffiltext{4}{Kavli Institute for the Physics and Mathematics of the Universe (Kavli IPMU, WPI), University of Tokyo, Chiba 277-8583}
\altaffiltext{5}{National Astronomical Observatory of Japan, Osawa, Mitaka, Tokyo, 181-8588}
\altaffiltext{6}{SOKENDAI (The Graduate University for Advanced Studies), Osawa, Mitaka, Tokyo 181-8588}



\maketitle

\begin{abstract}
We report a detailed modeling of a mass profile of a $z = 0.2999$ massive elliptical galaxy using 30 milli-arcsecond resolution 1-mm Atacama Large Millimeter/submillimeter Array (ALMA) images of the galaxy--galaxy lensing system SDP.81.  The detailed morphology of the lensed multiple images of the $z = 3.042$ infrared-luminous galaxy, which is found to consist of tens of $\lesssim 100$-pc-sized star-forming clumps embedded in a $\sim 2$~kpc disk, are well reproduced by a lensing galaxy modeled by an isothermal ellipsoid with a 400~pc core.  The core radius is consistent with that of the visible stellar light, and the mass-to-light ratio of $\sim 2\,M_{\Sol}\,L_{\Sol}^{-1}$ is comparable to the locally measured value, suggesting that the inner 1 kpc region is dominated by luminous matter.  The position of the predicted mass centroid is consistent to within $\simeq 30$~mas with a non-thermal source detected with ALMA, which likely traces an active galactic nucleus of the foreground elliptical galaxy.  While the black hole mass and the core radius of the elliptical galaxy are degenerate, a point source mass of $> 3 \times 10^8\,M_{\Sol}$ mimicking a supermassive black hole is required to explain the non-detection of a central image of the background galaxy.  The required mass is consistent with the prediction from the well-known correlation between black hole mass and host velocity dispersion.  Our analysis demonstrates the power of high resolution imaging of strong gravitational lensing for studying the innermost mass profile and the central supermassive black hole of distant elliptical galaxies.
\end{abstract}



\section{Introduction}

Galaxy--galaxy lensing occurs ubiquitously in the observable universe.  Magnified images from gravitational lensing allow us to understand the detailed mass structure of the intervening galaxy as well as the background source structure (e.g., \cite{Koopmans05, Inoue05, Vegetti12, Hezaveh13a, Hezaveh13b, Hezaveh14}).  Furthermore, a detection or absence of a central lensed image offers a unique opportunity to map the innermost mass distribution of the lensing galaxy and to directly measure the mass of  the central supermassive black hole (SMBH, \cite{Mao01, Winn04, Inada08}).  Precise mass modeling requires sensitive, extinction-free, and extremely high resolution imaging toward the very center of a lensing galaxy \citep{Hezaveh15}.  This is now possible with long baseline capabilities offered by the Atacama Large Millimeter/submillimeter Array (ALMA).   Here we present a detailed modeling of a mass profile of the $z = 0.2999$ elliptical galaxy, which is the foreground galaxy in the galaxy--galaxy lensing system H-ATLAS J090311.6+003906 (SDP.81), using 30 milli-arcsec (mas) resolution 1-mm images obtained at ALMA.

SDP.81 is a gravitationally lensed $z=3.042$ galaxy discovered in the \textit{Herschel} Astrophysical Terahertz Large Area Survey (H-ATLAS, \cite{Eales10}).  The characteristic features (``Einstein ring'') are evident in the $2''$ resolution interferometric maps \citep{Negrello10, Omont13, Bussmann13} and the \textit{Hubble Space Telescope} (HST) image \citep{Negrello14, Dye14}, indicating strong magnification.  Carbon Monoxide (CO) and water (H$_2$O) emission were successfully detected from the background galaxy \citep{Omont13}.  The foreground elliptical galaxy is identified in the Sloan Digital Sky Survey (SDSS) as SDSS~J090311.57+003906.5 (hereafter SDSS~J0903) with a spectroscopic redshift of $z = 0.2999 \pm 0.0002$ \citep{Negrello14}.  We hereafter refer to the foreground and background galaxies as SDSS~J0903 and SDP.81, respectively.  Detailed discussion on the spatial/kinematic structure and the star-formation activity in SDP.81 will be presented in a subsequent paper (Paper II; \cite{Hatsukade15}).

Throughout this paper, we assume a flat universe with $\Omega_\mathrm{m} = 0.26$, $\Omega_{\Lambda} = 0.74$, and $H_0 = 72$ km~s$^{-1}$~Mpc$^{-1}$. The angular scale of $1''$ corresponds to 4.36~kpc at the lens redshift $z_\mathrm{l} = 0.2999$ and 7.78~kpc at source redshift $z_\mathrm{s} = 3.042$.


\begin{table*}
\tbl{
Lens model parameters.  
}{%
\begin{tabular}{cccccccccccc}
\hline
$x$$^{\rm a}$ & 
$y$$^{\rm a}$ & 
$\sigma_{v}$$^{\rm b}$ & 
$e$$^{\rm c}$ & 
$\theta_{e}$$^{\rm c}$ & 
$r_\mathrm{core}$$^{\rm d}$ &
$\gamma$ ($\times 10^{-2}$)$^{\rm e}$ & 
$\theta_{\gamma}$$^{\rm e}$ & 
$\delta$ ($\times 10^{-3}$)$^{\rm f}$ & 
$\theta_{\delta}$$^{\rm f}$ & 
$\epsilon$ ($\times 10^{-3}$)$^{\rm g}$ & 
$\theta_{\epsilon}$$^{\rm g}$\\ 
\hline
$6 ^{+21}_{-32}$ & 
$5 ^{+13}_{-22}$ & 
$265 ^{+15}_{-4}$ & 
$0.14 ^{+0.11}_{-0.07}$ & 
$11 \pm 13$ & 
$48 ^{+169}_{-28}$ &
$4.0 ^{+1.7}_{-3.2}$ & 
$3 ^{+35}_{-23}$ & 
$3.0 ^{+3.1}_{-2.3}$ & 
$41 \pm 22$ & 
$2.4 ^{+3.7}_{-1.7}$ & 
$35^{+23}_{-13}$ \\
\hline
\end{tabular}}\label{tab:tab1}
\begin{tabnote}
The errors represent 68\% 
confidence intervals estimated from a Markov-chain Monte Carlo simulation.  
(a) The central position of an isothermal ellipsoid in units of mas, relative to the position of the non-thermal source at \timeform{9h3m11.573s}, \timeform{+0D39'6.54''}, respectively.
(b) Velocity dispersion of the isothermal ellipsoid (km s$^{-1}$).
(c) Ellipticity and its position angle (deg).
(d) core radius of the isothermal ellipsoid (mas).
(e) external shear and its position angle (deg).
(f) third-order external perturbation and its position angle (deg).
(g) multipole perturbation with $(m,\,n) = (4,\,2)$ and its position angle (deg). See \citet{Oguri10} for the definitions of the parameters.
\end{tabnote}
\end{table*}



\begin{figure*}
 \begin{center}
  \includegraphics[width = 16cm]{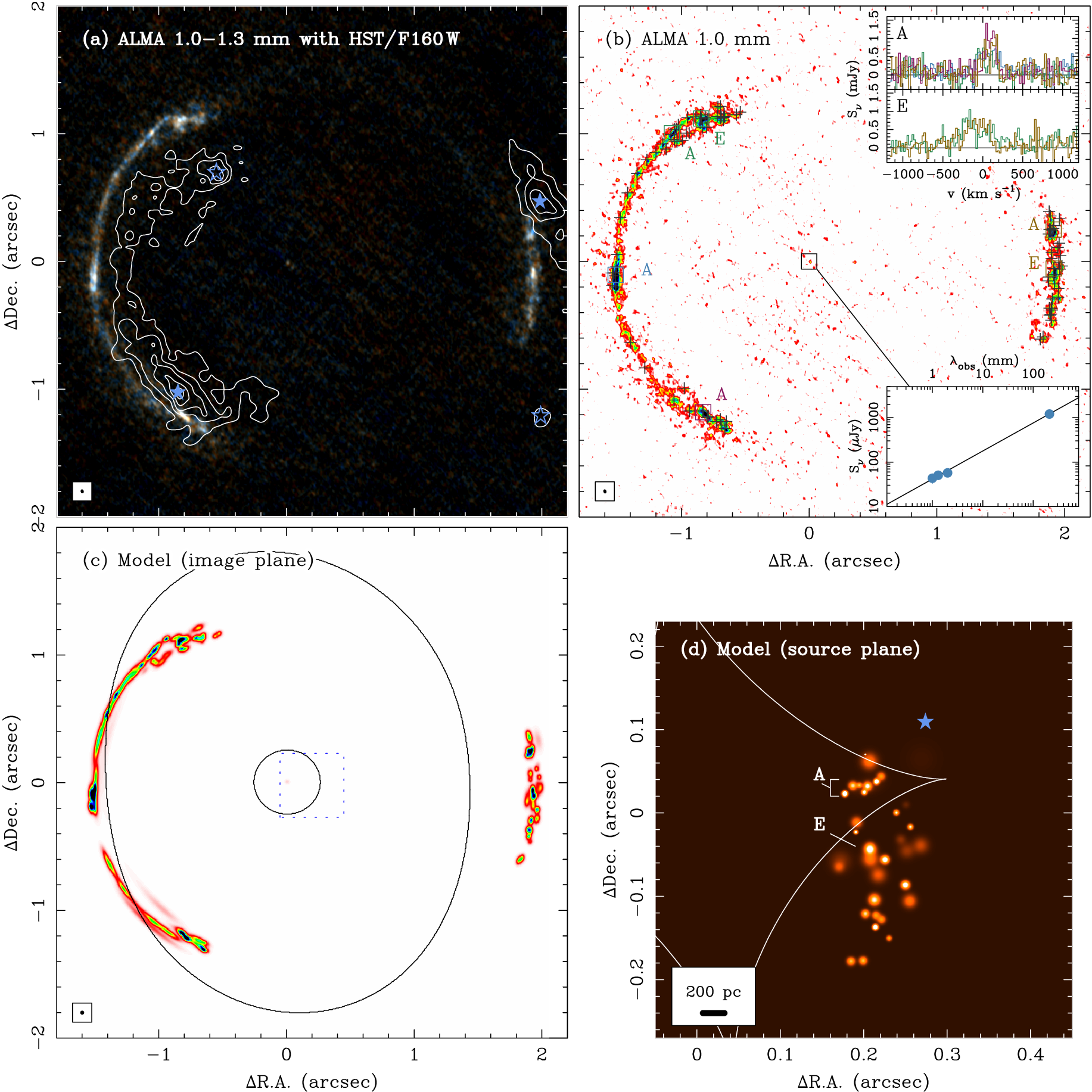}
 \end{center}
\caption{
(a) ALMA 3-color image of SDP.81 (1.0, 1.15 and 1.3~mm for blue, green and red, respectively) overlaid with the Hubble WFC3/F160W (1.6~$\micron$) image where the stellar light of SDSS~J0903 is subtracted (contours).  Two sets of counter-images of stellar peaks are indicated by filled and open stars, respectively.  The synthesized beam size is indicated at the bottom-left corner.  The origin of the image is taken at the position of a central compact non-thermal source.
(b) The ALMA 1.0~mm image. Upper inset shows CO (5--4) spectra at the positions of the source A (upper) and E (lower).  Bottom inset shows the spectral energy distribution of the central compact source, which is well fitted by a power-law function with a spectral index of $-0.64$ (solid line), suggesting the synchrotron emission.
(c) The modeled brightness distribution on the image plane.  The image is smoothed by a Gaussian with FWHM = 23 mas. The inner and outer ellipses represent radial and tangential critical curves, respectively.
(d) The modeled brightness distribution on the source plane, which is $0.5''$ on a side.  The star represents the source position of the stellar peaks denoted as filled stars in (a).  The position of this panel is indicated as a dotted square in (c). The solid curves represent the caustics.  The scale bar at the bottom-left corner shows a physical scale of 200~pc at $z = 3.042$.
}\label{fig:fig1}
\end{figure*}


\begin{figure*}
 \begin{center}
  \includegraphics[width=15.5cm]{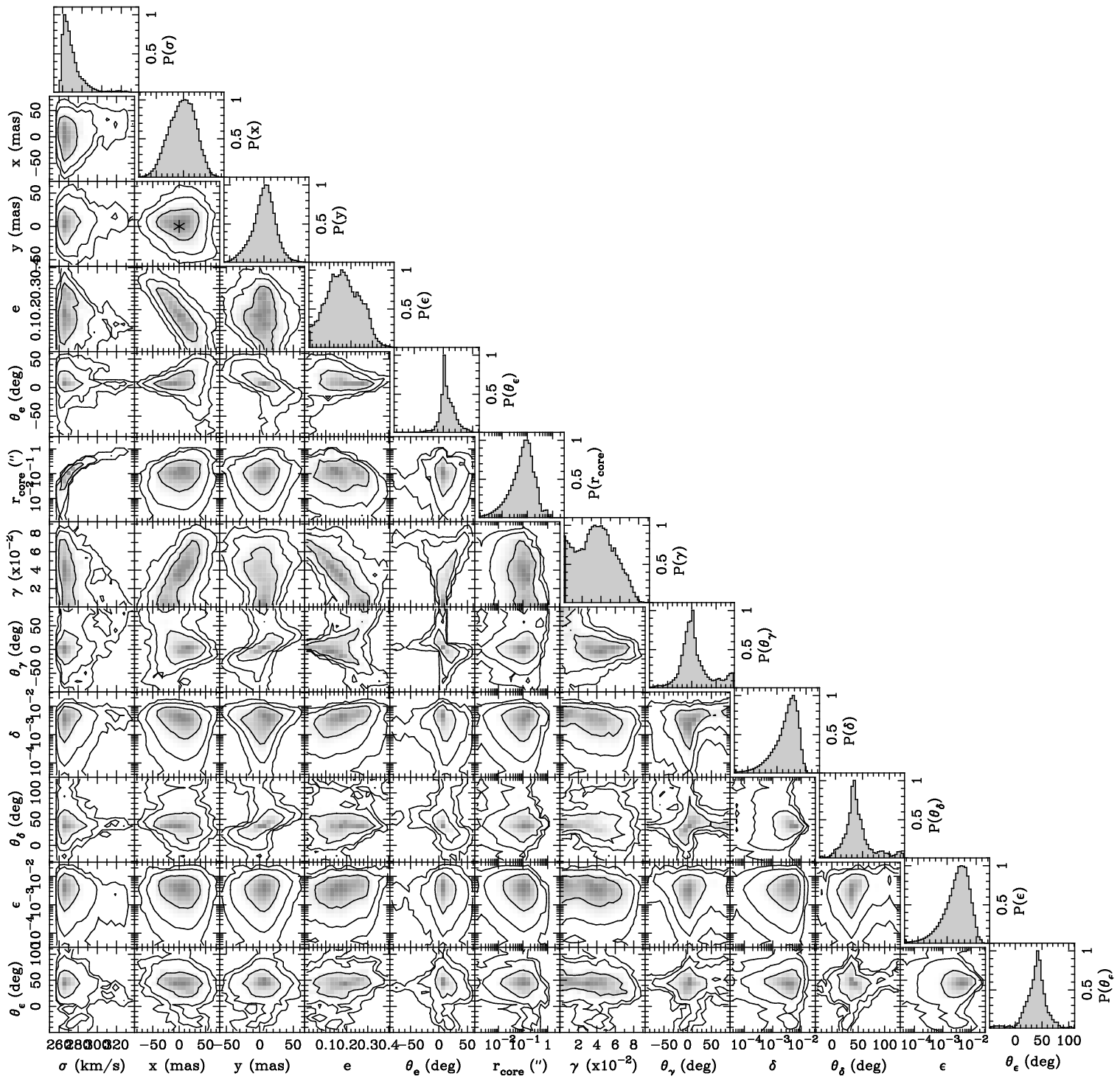}
 \end{center}
\caption{
Correlation matrix of the lens parameters computed by a Marcov-chain Monte Carlo simulation. The probability distribution of two parameters are indicated by a greyscale map with contours of $1\sigma$, $2\sigma$ and $3\sigma$. The diagonal panels show the probability distribution of each parameter in which the rest parameters are marginalized over.  The asterisk in the $x$--$y$ correlation plot marks the position of the AGN at the center of SDSS~J0903.
}\label{fig:fig2}
\end{figure*}


\section{Data}

\subsection{ALMA observations and images}

The ALMA data at wavelengths of 1.0, 1.3 and 2.0~mm were obtained during the long baseline campaign of Science Verification \citep{Fomalont15}.  The resulting spatial resolution achieved at 1.0, 1.3 and 2.0~mm are $23 \times 31$ mas (PA = 15$\arcdeg$), $30 \times 39$ mas (PA = 20$\arcdeg$) and $54 \times 60$ mas (PA = 18$\arcdeg$), with r.m.s.\ noise level of 10, 10, and 9~$\mu$Jy beam$^{-1}$, respectively.  The CO (5--4), (8--7) and (10--9) images were also taken simultaneously in the 2.0, 1.3 and 1.0~mm bands, respectively.  The CO images are $uv$-tapered to $\simeq 170$~mas in order to increase the sensitivity to extended structure.  Details of the observations are given in \citet{Vlahkis15} and \citet{Fomalont15}.

A three-color image composed from the 1.0 and 1.3~mm data are shown in figure~\ref{fig:fig1}a. The quadruple images of SDP.81 are clearly resolved, three of which are bridged by a $3''$-long arc.  The arcs consist of four bright knots connected by thin smooth filaments.  The grainy clumps seen around the arcs are not coherent among the different frequency bands (figure~\ref{fig:fig1}a), and thus they are likely to be artifacts due to phase calibration residuals.  On the other hand, we find fainter, but significant knots embedded in the arcs that are consistent in all different waveband images.  These are real features that are likely tracing the internal substructures of SDP.81.  
A central point source bracketed by the long and short arcs is not resolved with the Band 7 beam and has a 1.0--2.0~mm power-law spectral index of $\alpha = -0.3 \pm 0.7$ \citep{Vlahkis15}.  The extrapolated flux at 20~cm is consistent with that of the cataloged radio source FIRST~J090311.5+003906 ($S_\mathrm{20cm} = 1.21 \pm 0.16$~mJy~beam$^{-1}$, \cite{Becker95}).  The power-law spectral index from 1.0~mm to 20~cm is $\alpha = -0.64 \pm 0.01$ (see inset of figure~\ref{fig:fig1}b), which is consistent with the spectral index measured for 1.0--2.0 mm.  If the non-thermal emission arises from SDP.81, it should also dominate the emission from the arcs.  However, the observed spectral index of the arc is positive (typically $\alpha \gtrsim 2$), meaning that dust emission is the better model to explain the emission characteristics of the arc \citet{Vlahkis15}.  Furthermore, no molecular lines of CO and H$_2$O at $z=3.042$ are found at the position.  It is therefore likely that the non-thermal emission arises from an Active Galactic Nucleus (AGN) of SDSS~J0903.


\subsection{Hubble images}

We retrieve the archival HST image (proposal ID: 12194, \cite{Negrello14}) taken with the Wide Field Camera 3 (WFC3) at 1.6~$\micron$ (F160W, \textit{H} band).  The point spread function (PSF) is 151~mas in FWHM.  We calibrate the coordinates of the WFC3 image using the catalog data from the SDSS Data Release 12\footnote{http://skyserver.sdss.org/dr12}. We use 46 SDSS objects located in the WFC3 image to derive the astrometric fit using \textsc{iraf}.  The r.m.s.\ errors of the fit are 85~mas in right ascension and 100~mas in declination.  The nominal center of SDSS~J0903 matches the position of the central non-thermal source in the ALMA images within 15~mas.

We use the \textsc{galfit} software (Version 3.0.5, \cite{Peng10}) to deblend the stellar light in SDP.81 from SDSS~J0903.  Following the procedure adopted by \citet{Negrello14}, two S\'{e}rsic profiles are simultaneously fitted to the HST image.  As a result, the radial profile of SDSS~J0903 is decomposed into a brighter (\textit{H}-band absolute AB magnitude of $M_H = 17.56$) elliptical with a S\'{e}rsic index of $n_\mathrm{s} = 4.42$ and a fainter ($M_H = 18.69$) exponential disk with $n_\mathrm{s} = 1.03$.  Although our total flux is $1.2\times$ higher, this is consistent with the model presented by \citet{Negrello14}.  The residual image shown in figure~\ref{fig:fig1} clearly exhibits the stellar arcs which are significantly ($\gtrsim 100$~mas) offset from the 1-mm ALMA arcs (see Paper II for the interpretation).


\begin{figure}[t]
 \begin{center}
  \includegraphics[angle=-90, width=8cm]{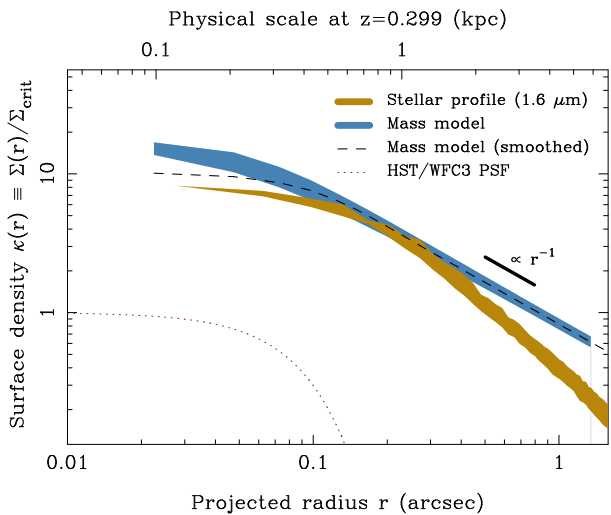}
 \end{center}
\caption{
Surface mass density versus projected radius of the best-fit mass model (blue shade), radial distribution of stellar light (dark yellow) and the point spread function (PSF) of the 1.6 $\micron$ HST image (dotted curve, arbitrary unit).  The mass model smoothed by the HST PSF is shown in dashed curve.  The surface density is normalized by the cosmological critical density for a source redshift $z_\mathrm{s} = 3.042$, $\Sigma_\mathrm{crit} = 4.3 \times 10^9$ $M_{\solar}$ kpc$^{-2}$.
}\label{fig:fig3}
\end{figure}



\section{Gravitational lens model}

\subsection{Methods and results}

We use the \textsc{glafic} software \citep{Oguri10}\footnote{http://www.slac.stanford.edu/\%7Eoguri/glafic/} 
to reproduce the lensed image using an isothermal ellipsoid with a flat core.  The choice of an isothermal ellipsoid is supported by the fact that the radial density profile of SDSS~J0903 is consistent with $\rho(r) \propto r^{-2}$ \citep{Dye14, Rybak15}, which is the profile expected for an isothermal ellipsoid.  The structure of SDP.81 is, however, too complex to be accounted for by a single isothermal ellipsoid alone.  We examine the SDSS cluster catalog of \citet{Oguri14} and find a cluster of galaxies at $z = 0.306$ with the richness of $\sim 21$ and located $\sim 2'$ away from SDSS~J0903.  Therefore we include the third-order and quadrupole perturbations of mass\footnote{We use \textsc{glafic} mass models \texttt{clus3} for the third-order external perturbation and \texttt{mpole} with $(m, n) = (4, 2)$ for the quadrupole perturbation (equation~13 of \cite{Oguri10}).  The central positions of the perturbation components are fixed to that of the isothermal ellipsoid.}, in addition to the standard external shear, to take into account the gravitational influence of the cluster and large-scale structure.

We identify compact knots in the arcs that are common in the 1.0, 1.3, and 2.0~mm images and then identify two sets of counter-images of the knots in SDP.81.  These are shown as open squares in figure~\ref{fig:fig1}b (A and E). The counter-images of each set are extracted by visually identifying continuum intensity peaks which share the same/similar CO (5--4) and (8--7) spectral profiles at each of the peaks.  Figure~\ref{fig:fig1}b shows how the counter-images are identified; all of the CO (5--4) spectra at the positions of A have a peak at $\sim 100$ km~s$^{-1}$ while the peaks at E appear near $-200$ km~s$^{-1}$ and have broader line profiles than A, which likely correspond to the red and blue components reported by \citet{Omont13} and \citet{Vlahkis15}, respectively.  This method allows us to predict that the knots A are quadruple lensed images of a single source, whereas the knots E are double images and thus the (tangential) caustics should lie between A and E on the source plane.  First, we obtain an initial mass model only with the isothermal ellipsoid and external shear using the positions of A and E.  Then we use the initial mass model to identify groups of counter-images of knots found in the 1.0-mm continuum image.  The knots within $\simeq 50$ mas from the predicted positions are identified as the counter-images, resulting in identifications of 46 images.  We then refine the mass model by adding the higher-order perturbations (the number of independent parameters is 12) and fitting all of the identified multiple images. In the model calculation, we also consider two sets of double images identified with the 1.6-$\micron$ HST image, which is indicated by star symbols in figure~\ref{fig:fig1}a.  The number of degrees of freedom (DOF) is thus $N_\mathrm{DOF} = 88$.  The positional uncertainty of the knots used for optimization is set to 20~mas, while the uncertainties of the 10 knots which are close to the critical curve and/or possibly suffering from blending are set to 100--200~mas.  The uncertainties of the four HST images are set to 130--200~mas.

The best-fit parameters of the mass model\footnote{The \textsc{glafic} model of SDP.81 is available at http://www.ioa.s.u-tokyo.ac.jp/\%7Eytamura/SDP81/.} 
are listed in table~\ref{tab:tab1}, where the $1\sigma$ confidence intervals are computed by Marcov-chain Monte Carlo simulations as shown in figure~\ref{fig:fig2}.  Note that the core radius ($r_\mathrm{core}$) and velocity dispersion ($\sigma_v$) of the isothermal ellipsoid are highly degenerate.  The central position of the isothermal ellipsoid is not only consistent with the brightness peak of the stellar distribution within the $1\sigma$ astrometric uncertainty of HST ($\simeq 100$~mas), but also with the location of the candidate AGN, i.e., $(x,\,y) = (0,\,0)$, to within $\simeq 30$~mas ($1\sigma$).  This suggests that the AGN is indeed located at the bottom (innermost $< 100$-pc) of the potential well of SDSS~J0903.  The best-fit parameters of the isothermal ellipsoid and external shear are in good agreement with those independently obtained using the same ALMA images \citep{Rybak15,Dye15}\footnote{Note that the ellipticity $q$ used in \citet{Rybak15} is related by $e = 1 - q$.} and the HST images \citep{Dye14} at $1\sigma$ level.

The crosses in figure~\ref{fig:fig1}b mark the positions of the counter-images predicted by this model, and we find that the model is qualitatively consistent with the peak positions of the 1.0-mm brightness distribution ($\chi^2 = 3.45$).  Figures~\ref{fig:fig1}c and \ref{fig:fig1}d show the predicted brightness distributions on the image and source planes, respectively.  The observed images are reproduced by 35 $\lesssim 100$-pc-sized knots embedded in an extended ($\simeq 2$~kpc) disk.  While the uncertainty is relatively large, the total magnification factor is expected to be $\mu_\mathrm{g} \sim 22$. This is consistent with the value obtained by \citet{Rybak15} ($\mu_\mathrm{g} = 17.6 \pm 0.4$ for total, $\mu_\mathrm{g} = 25.2 \pm 2.6$ for a central star-forming disk), while it is $\simeq 2 \times$ higher than the value previously obtained with the SMA ($\mu_\mathrm{g} \simeq 11$, \cite{Bussmann13}).  This is likely because, as suggested by \citet{Rybak15}, the ALMA image allows us to identify more compact structures close to the caustics, resulting in higher magnifications than those obtained previously.


\subsection{Mass profile of SDSS J0903}

In figure~\ref{fig:fig3}, we show the surface mass profile of the gravitational lens model normalized by a cosmological critical density for a source redshift $z_\mathrm{s} = 3.042$ (i.e., convergence $\kappa(r)$) and a radial profile of the 1.6-$\micron$ stellar light obtained by the HST/WFC3 F160W imaging observations.  The critical density is $\Sigma_\mathrm{crit} = 4.3 \times 10^9$ $M_{\solar}$ kpc$^{-2}$, and the stellar surface mass is scaled using a fiducial mass-to-light ratio of $2~M_{\solar}~L_{\solar}^{-1}$ at $\lambda_\mathrm{rest} \simeq 1~\micron$ (e.g., \cite{Bell03}), which is a typical value for local massive ellipticals.  Because the stellar profile is lowpass-filtered by the PSF of HST (151~mas, see the dotted curve), the profile is only reliable at $r \gtrsim 0.1''$.  It is evident that there is a knee in the stellar profile at a characteristic radius of $\simeq 0.15''$.

In figure~\ref{fig:fig3}, we also show the lensing mass profile smoothed by the HST PSF.  The knee position and the amplitude of the profile are in agreement with the lens model.  Although the derived core ($r_\mathrm{core} = 48 ^{+169}_{-28}$~mas, which corresponds to a core diameter of $\sim$400~pc) is slightly smaller than the stellar core ($r \simeq 150$~mas), the two core radii are consistent to within the $1\sigma$ uncertainty.  Furthermore, the position angle of the isothermal ellipsoid ($\theta_e = 11 \pm 13 \arcdeg$) is consistent with that of the stellar profile ($10 \pm 1\arcdeg$, \cite{Dye14}), suggesting that the baryonic matter, traced by the stellar light, dominates the mass in the inner 1-kpc region of the galaxy.  At larger radii ($\gtrsim 3$~kpc), however, the stellar light falls below the model surface density ($\kappa(r) \propto r^{-1}$), which can be interpreted as the presence of dark matter in the outer parts of SDSS~J0903.


\begin{figure}[t]
 \begin{center}
  \includegraphics[angle=-90,width=8cm]{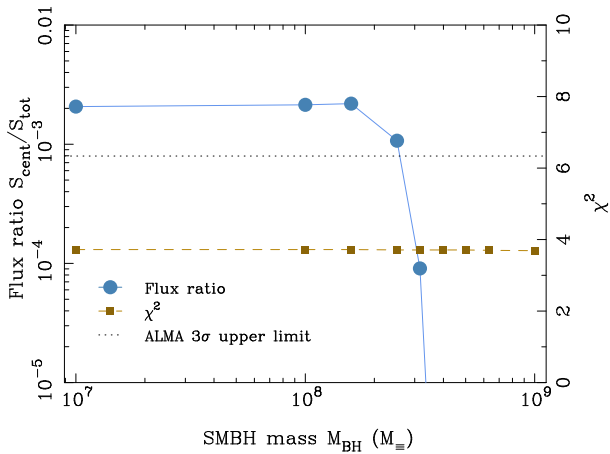}
 \end{center}
\caption{
The flux density ratio of central counter-images relative to the all counter-images, as a function of mass of a supermassive black hole of SDSS~J0903 (Solid line with circles). The source-plane components are the same as shown in figure~\ref{fig:fig1}d.  The horizontal dotted line shows the $3\sigma$ upper limit of the ALMA 1.0-mm image on the flux ratio.  The dashed line with squares represents $\chi^2$ of model optimizations. 
}\label{fig:fig4}
\end{figure}


\section{Discussions and conclusions}

The non-detection of the central `odd' image of SDP.81 in the deep ALMA images rules out a top-flat plateau in the gravitational potential, but instead requires the presence of a steep (or cuspy) radial profile in the innermost potential distribution.  A SMBH can make the central potential more cuspy even if the stellar mass distribution is top-flat as suggested from the WFC3 radial profile. As a consequence, this can erase the central image if the black hole is massive enough to fully perturb the core potential.  Since there is a clear sign of the presence of an AGN at the center of SDSS~J0903, we add a point source mimicking the SMBH to the lens model in order to investigate how the SMBH affects the flux of the central image.  A point mass with $M_\mathrm{BH} = 10^7$ to $10^9\,M_{\Sol}$ is placed at the position of the AGN and $r_\mathrm{core}$ is fixed to the stellar core radius of $150$~mas, while the rest are treated as free parameters ($N_\mathrm{DOF} = 89$).

Figure~\ref{fig:fig4} shows S$_{\rm cent}$/S$_{\rm tot}$ (i.e. the ratio between the flux from the central image to all of the counter-images) as a function of $M_\mathrm{BH}$. The fit to the model is generally good ($\chi^2 \simeq 3.7$) in the range of $10^7 < M_\mathrm{BH}/M_{\Sol} < 10^9$. The ratio is insensitive to the black hole mass if it is less than $1 \times 10^8$ $M_{\Sol}$, but the central image rapidly dims as the black hole mass is higher than $2 \times 10^8 M_{\Sol}$.  The horizontal dotted line shows the $3\sigma$ upper limit of the 1.0-mm image (30~$\mu$Jy) divided by the total flux ($37.7 \pm 1.5$ mJy, \cite{Vlahkis15}), suggesting that the ALMA non-detection places a constraint to the black hole mass of $M_\mathrm{BH} > 3 \times 10^8\,M_{\Sol}$.  Note that if we adopt $r_\mathrm{core} = 48$~mas, the limit is slightly relaxed to $M_\mathrm{BH} > 1 \times 10^8\,M_{\Sol}$.

A black hole mass can independently be estimated from the well-known correlation between black hole mass and the velocity dispersion of the bulge (e.g., \cite{Ferrarese00, Gebhardt00}). The black hole estimated from the velocity dispersion of the lens ($\sigma_v = 265 ^{+15}_{-4}$~km~s$^{-1}$, Table~\ref{tab:tab1}) and using the local correlation \citep{Kormendy13} is $M_\mathrm{BH} = (1 \pm 0.1) \times 10^9$ $M_{\Sol}$, which is consistent with the lower limit derived above.  It will difficult to detect the central image of SDP.81 even in future ALMA observations if $M_\mathrm{BH} \sim 1 \times 10^9\,M_{\Sol}$.  We demonstrate in this work that the detailed gravitational lens modeling using sensitive, high-resolution imaging with ALMA is capable of constraining the central $\lesssim 1$~kpc mass profile and the black hole mass, which will help us understand the co-evolution of the SMBHs and the host galaxies through, for example, SMBH scouring \citep{Hezaveh15}.

Finally, we note that a similar conclusion on the mass of the SMBH ($\log{M_\mathrm{BH}/M_{\Sol}} > 8.4$) was made by a recent analysis by \citet{Wong15}, where they used an independent code \textsc{glee} \citep{Suyu10} and different sets of multiple images from the ALMA images tapered to $\sim 170$ mas to construct a lens model.  Nevertheless, we prefer to use the full resolution ALMA images (up to 23 mas), which allows us to (1) accurately model higher-order perturbations, and to (2) robustly identify multiple features in the images and construct a reliable lens model.


\begin{ack}
We thank the anonymous referee for fruitful comments.  This work is supported by KAKENHI (No.\ 25103503, 26800093 and 20647268). MO is supported in part by World Premier International Research Center Initiative (WPI Initiative), MEXT, Japan.  
DI is supported by the 2015 Inamori Research Grants Program.
This paper makes use of the following ALMA data: ADS/JAO.ALMA \#2011.0.00016.SV.  ALMA is a partnership of ESO (representing its member states), NSF (USA) and NINS (Japan), together with NRC (Canada), NSC, ASIAA (Taiwan) and KASI (Republic of Korea), in cooperation with the Republic of Chile. The Joint ALMA Observatory is operated by ESO, NAOJ and NRAO.
\end{ack}





\end{document}